# Three dimensionality of pulsed second-sound waves in He II

P. Zhang[1] and M. Murakami[2]

[1] Institute of Refrigeration and Cryogenics, Shanghai Jiao Tong University, Shanghai, 200030, China

[2] Graduate School of Systems and Information Engineering, University of Tsukuba, Tsukuba, 305-8573, Japan

ABSTRACT

Three dimensionality of 3D pulsed second sound wave in He II emitted from a finite size heater is experimentally investigated and theoretically studied based on two-fluid model in this study. The detailed propagation of 3D pulsed second sound wave is presented and reasonable agreement between the experimental and theoretical results is obtained. Heater size has a big influence on the profile of 3D second sound wave. The counterflow between the superfluid and normal fluid components becomes *inverse* in the rarefaction of 3D second sound wave. The amplitude of rarefaction decreases due to the interaction between second sound wave and quantized vortices, which explains the experimental results about second sound wave near $T_\lambda$ [Phys. Rev. Lett. **73**, 2480 (1994)]. The accumulation of dense quantized vortices in the vicinity of heater surface leads to the formation of a thermal boundary layer, and further increase of heating duration results in the occurrence of boiling phenomena.

I. INTRODUCTION

Superfluid helium (He II) is an ideal system which can be employed to study nonlinear acoustics since He II holds very high purity and other impurities are condensed at low temperature environment. He II is composed of two components: superfluid and normal fluid, according to Landau's two-fluid model [1], which leads to the existence of a temperature wave-second sound wave in addition to a normal pressure wave-first sound wave. Second sound wave is generated by the counterflow between the superfluid and normal fluid components induced normally by a heat current emitted by a heater (second sound wave emitter) immersed in He II. Second sound wave transmits as planar wave in 1D geometry which can be observed in a channel with one end perfectly bounded by a heater or in an unbounded He II bath with an infinite size heater. For large amplitude planar second sound wave, the transmitting velocity becomes amplitude dependent. The first experimental observation of nonlinear second sound wave was reported by Osborne [2]. Nonlinear theory about second sound wave was first developed by Temperley [3], and was then extended by Khalatnikov [4, 5] whose prediction was proven by the experimental demonstration of Dessler and Fairbank [6]. Planar second sound wave has been frequently used to probe quantized vortices for its relatively simple geometry, and extensive study concerning this topic has been carried out in recent decades [7-10]. It has been shown that second sound wave interacts with self-generated quantized vortices as it propagates along a wave guide. For small heat current, the shape of second sound wave remains the same as that of the input heat pulse and heat can be entirely transported by second sound wave which is almost free of quantized vortices; however, for large heat current, quantized vortices are generated due to larger relative velocity $v_{ns} = |v_n - v_s|$ between the superfluid and normal fluid components than a certain critical value $v_c$ and second sound wave develops into limiting profile [11].

In many cases, heater is of finite size, which generally results in a non-planar second sound wave. Previous study of second sound wave emitted from a finite-size heater has shown that non-planar second sound wave is composed of a positive temperature excursion followed by a negative temperature rarefaction [12]. The first comprehensive experimental study of non-planar second sound wave might have been carried out by Iznankin and Mezhov-Deglin [13], and non-planar first sound wave was also measured at the same time; further theoretical explanation of non-planar second sound wave was based on the interpretation by introducing a hydrodynamic potential via which the problem was described by wave equation similar to that used for non-planar first sound wave. Later, similar approach for the interpretation of non-planar second sound wave was also adopted by Efimov, et al. [14]. Some other theoretical considerations [12] of non-planar second sound wave generally started from the modified hyperbolic heat transfer equation, and it seems that these approaches did not reflect the superfluid nature at all, although they could produce a wave-like result in temperature excursion. Krysac [15] measured non-planar second sound wave near $T_\lambda$ and presented a vector representation method by assuming a convection field towards or away from the heater depending on the sign of the temperature amplitude and integrating around the heater. Nevertheless, the existence of such an assumed convection filed seems unreasonable and moreover this method can only give out the qualitative results. It seems that the propagation process of non-planar second sound wave is not fully understood yet and more reasonable interpretation is required. This paper presents the first numerical study of the propagation of 3D pulsed second sound wave in He II from the starting point of two-fluid model. Changing from 1D to 3D second sound wave is numerically shown by varying the heater size. The detailed propagation process of non-planar second sound wave can be understood from the calculation results which are also validated by the experimental measurement.

II. THEORETICAL CONSIDERATION AND EXPERIMENTAL DESCRIPTION

The description of hydrodynamic behavior of He II by two-fluid model can be formulated as

$$\begin{aligned}
&\frac{\partial \rho}{\partial t} + \nabla \cdot (\rho v) = 0 \\
&\frac{\partial (\rho s)}{\partial t} + \nabla \cdot (\rho s v_n) = F_{ns} v_{ns} / T \\
&\frac{\partial v_s}{\partial t} + (v_s \cdot \nabla) v_s + \nabla \mu = F_{ns} / \rho_s \\
&\frac{\partial (\rho v_i)}{\partial t} + \frac{\partial \Pi_{ik}}{\partial x_k} = 0
\end{aligned} \quad (1)$$

where $\rho, s, v$ and $T$ are the density, the entropy, the velocity and the temperature. Subscripts $n$ and $s$ indicate the normal fluid and superfluid components, respectively. $\mu$ is the chemical energy defined as: $d\mu = -sdT + \frac{1}{\rho}dp - \frac{1}{2}\frac{\rho_n}{\rho}dv_{ns}^2$. $v_{ns}$ and $v_c$ are the relative and critical velocities, $F_{ns} = \frac{\kappa}{3}\frac{\rho_s \rho_n}{\rho} B_L L v_{ns}$ is the mutual friction force, $\kappa$ is the quantum of superfluid circulation, $B_L$ is coefficient and $L$ is the vortex line density (i.e. length of the vortex line in the unit volume). The above equations are the equations of continuity, conservation of entropy, motion of the superfluid component and conservation of the total momentum. The non-zero terms at the right side, representing the entropy production term and the mutual friction force term, are included to take the effect of quantized vortices into account. The form of these terms was first proposed by Vinen [16]. $\Pi_{ik}$ is the momentum flux density tensor for He II flow field, defined as

$$\Pi_{ik} = p\delta_{ik} + \rho_n v_{ni} v_{nk} + \rho_s v_{si} v_{sk} \qquad (2)$$

Second sound wave attenuates due to the interaction with quantized vortices resulting in the mutual friction between the superfluid and normal fluid components during the propagation. The description of quantized vortices through VLD, $L$, was first formulated by Vinen [16] and was further modified by Nemirovskii [17, 18] to add the field property of quantized vortices. The VLD equation is finally formulated as

$$\frac{\partial L}{\partial t} + \nabla \cdot (v_L L) = \frac{\chi_1 B_L \rho_n}{2\rho} v_{ns} L^{3/2} - \frac{\chi_2 h}{2\pi m} L^2 + \gamma |v_{ns}|^{5/2} \qquad (3)$$

where $v_L$ is the drift velocity of quantized vortices tangle, which is in the order of the velocity of the superfluid component, $v_s$ [19]. Three terms at the right side represent the development, decay and the source of quantized vortices, respectively.

The total set of two-fluid equations with Vinen's VLD equation, i.e. eqs. (1) and (3), are quite complicated and in highly nonlinear form. The theoretical analysis of these equations is very difficult even in 1D case and it seems that only numerical calculation is more feasible. There are many numerical studies [10, 20] of the propagation of 1D planar second sound wave, among which the approach by using MacCormack (Mac) scheme with Flux Corrected Transport (FCT) method can be extended to higher dimensional situation, i.e. two or three dimensions. The MacCormack two-step predictor and corrector method has second order accuracy in both time and space steps. FCT is used to suppress numerical oscillation and to preserve physical discontinuity of the wave front. The standard Mac-FCT method includes eight steps, the detail of which can be found in [21]. The determination of the thermodynamic quantities of He II is very crucial in the calculation. Unlike ordinary fluids, such as water, etc, the thermodynamic quantities of He II are not fully formulated, fortunately, they are well tabulated [22]. Therefore, thermodynamic quantities, $f(T,P)$, which are actually both temperature and pressure dependent, are determined by implementing the interpolation along both temperature and pressure axis. After the interpolation, the thermodynamic quantities are corrected by

$$\rho(P,T,v_{ns}) = \rho(P,T) + \frac{1}{2}\rho^2 v_{ns}^2 \frac{\partial}{\partial P}(\frac{\rho_n}{\rho}), \quad s(P,T,v_{ns}) = s(P,T) + \frac{1}{2}v_{ns}^2 \frac{\partial}{\partial T}(\frac{\rho_n}{\rho}) \text{ and } \mu(P,T,v_{ns}) = \mu(P,T) - \frac{1}{2}\frac{\rho_n}{\rho}v_{ns}^2$$

to account for their dependence on the relative velocity [1]. And the detailed description of the boundary conditions needed for the calculation is shown in Appendix A.

The calculation domain is one fourth of a cubical space formed by perfectly reflecting boundary walls, as shown in Figure 1, and He II is filled inside. The dimension is 45mm(AB) × 45mm(AD) × 90mm(AA'). The solid-filled area in the figure is one fourth of the heater placed in the center of the bottom (ABCD). If the bottom is totally bounded by the heater, the concerned problem is simplified to 1D. It is a 2D problem if the length of one side of the heater equals AD. The most complicated case is 3D problem shown in the figure. 2D slices of second sound wave shown in the paper are along the plane of ABB'A' if not specified. In the calculation, 3D pulsed second sound waves with the heater sizes of 25mm × 25mm, 45mm × 45mm and 65mm × 65mm in the case of heat flux of 10.0W/cm$^2$ and heating duration of 0.5ms and 1.0ms are studied for comparison. It should be noted that the pressure fluctuation in second sound wave has also been taken into account during calculation.

The experiment was to measure the evolution of the temperature excursion of second sound wave with time at a location in unperturbed He II bath. A thin film planar NiCr (about 60nm in thickness) heater of 25mm×25mm in size was employed as the second sound wave emitter. The typical resistance of the heater was about 30Ω, and the voltage applied to the heater was about 43V at a heat flux of about 10.0W/cm$^2$, which is generated by a wave form generator and is then amplified by a high speed precision amplifier. The temperature signal of second sound wave was measured by a superconductive temperature sensor, which was fabricated by vacuum deposition of gold and tin onto the outer surface of a 40 $\mu$m-diameter glass fiber. The temperature variation related to second sound wave, which induced a superconducting to normal-conducting transition of gold-tin film [23], resulted in the proportional change of the voltage drop across the glass fiber at a constant current. The transition temperature of the sensing element was adjusted to the present experimental temperature range by trimming the thickness ratio of gold and tin. The typical thickness of gold and tin was about 20nm and 100nm, respectively, and the dimensionless sensitivity of the sensor $S_d$ ( $S_d = d\ln(V)/d\ln(T)$ ), was larger than 3.0 at 1.6K, and it decreased to less than 1.0 at temperature close to $T_\lambda$. A single heat pulse was applied to the heater to generate pulsed second sound wave in He II. The waiting time between the next pulses is long enough for the existing quantized vortices to decay to the initial level. The sensor was placed right above the heater surface to measure the temperature excursion and it was calibrated against the saturated vapor pressure of He II before and after the measurement to guarantee its stability. The temperature signal of second sound wave measured by a superconductive temperature sensor was 100 times amplified by a high speed low noise pre-amplifier and was then recorded by the data recording system.

III. RESULTS AND DISCUSSION

A. 3D pulsed second sound wave and its propagation

Shown in Figures 2 (a) and (b) are the typical results of the temperature excursion of 3D pulsed second sound waves at He II bath temperature of 1.6 K with heating durations of 0.5ms and 1.0ms at heat flux of 10.0W/cm$^2$, the experimental results are taken at a distance $d_T$ =2.0mm above the heater surface along the central axis. It is seen that the calculation results agree quite well with the experimental ones. The characteristic feature of 3D pulsed second sound wave is that a rarefaction is found to follow the positive temperature excursion. The reason why the rarefaction appears has ever been proposed in [24], however, these explanations were based on the modified hyperbolic heat transfer equation or wave equation and it seems that they did not reflect any superfluid nature. The qualitative interpretation [13] has been proposed by using the similar procedure to that for first sound wave (pressure wave) [1]. As second sound wave expands outwards, the increase in the entropy flux which is proportional to $4\pi r^2 \rho_n (v_n - v_s)$ in the positive temperature excursion will have to be off-set by a decrease of the entropy flux in the region behind the positive second sound wave due to energy conservation. The positive temperature excursion gradually ends when the heating is shut off, and thus, the rarefaction appears.

From the starting point of two-fluid model with VLD equation, it is found from the numerical results that the rarefaction, i.e. negative temperature excursion, gradually appears in the vicinity of the heater surface after the heating is switch off. As can be seen from a series of 2D slices of 3D pulsed second sound wave at different time instances shown in Figures 3 and 4, rarefaction initially forms close to the edge of the heater (Figure 3 (b)), it gradually propagates inwards as the positive second sound wave portion further expands outwards and the rarefaction finally converges in the center of the heater surface (Figure 3 (c)). At this moment, the rarefaction has already well formed above the heater surface. After the rarefaction propagates away from the center of the heater surface, another compression wave with positive temperature amplitude (much weaker) is further observed there, as shown in Figure 3 (d). This weaker compression wave is induced by the expansion of the rarefaction outwards. Such consequent attenuated wave motion is also confirmed in another calculation case, as shown in Figure 4.

It is further seen from Figures 3 (d) and (e) that the counterflow between the superfluid and normal fluid components becomes *inverse* in the rarefaction portion [25]. The heat transport in He II results from the counterflow: superfluid component bearing no entropy flows towards the heat source, while normal fluid component flows away from the heat source to keep the mass conservation, yielding $q = \rho_s s T v_{ns}$ by assuming $\rho v = 0$ which is approximately true in the present case. In the positive temperature portion of the running second sound wave (free of reflection), $q > 0$ leads to $v_{ns} > 0$; and in the rarefaction portion, it seems quite natural that $q < 0$ leads to $v_{ns} < 0$, as shown by the relative velocity distribution in Figure 3 (e).

As can be seen from Figures 3 (a)-(d) and Figures 4 (c)-(e), second sound wave gradually diffracts away from the heater. It is obvious that the transmitting speed of second sound wave depends on its amplitude [4, 5]:

$$c_2 = c_{20}[1 + \Delta T \partial \ln(c_{20}^3 C_P / T) / \partial T] = c_{20}[1 + \varepsilon(T)\Delta T] \qquad (4)$$

where $c_{20}$ is the second sound wave velocity at zero amplitude, $\Delta T$ is the amplitude, $C_P$ is the specific heat capacity and $\varepsilon(T)$ is the nonlinear coefficient. The temperature amplitude of second sound wave along the central axis of the heater is larger than off-axis temperature amplitude, as seen from Figures 3 and 4, so that the transmitting speed along the central axis will be larger when $\varepsilon(T) > 0$, which may result in the gradual variation of the shape of the wave front. The shape of the wave front during propagation is like a spherical shell [26]. As the wave front expands, the temperature amplitude decreases and higher temperature range shrinks due to the nonlinearity feature of He II and energy conservation for 3D wave. The rarefaction also shows the similar propagating feature as the positive temperature portion does. Second sound wave reflects as it reaches the boundary wall of the container [27]. The temperature rise of the reflected portion is doubled, which results in the larger transmitting speed of second sound wave, and leads to the further change of the curvature of the wave front, as shown in Figure 3 (d) and Figures 4 (c)-(e).

It is obvious that the profile of second sound wave is an ideal 3D one in the case of the point heater. However, the shape of the profile is drastically influenced by the heater size when the heater is of finite size. Shown in Figure 4 (a) are the wave profiles of 3D pulsed second sound wave at $d_T$=2.0mm with different heater sizes. It is seen that the wave profile is bipolar type for the heater size of 25mm×25mm. As the heater size increases further to 65mm×65mm, a period of constant temperature equaling to He II bath temperature is observed after the initial positive temperature excursion and is followed by a rarefaction. The appearance of the rarefaction is somewhat delayed, which is quite counter-intuitive. As in the case of heater size of 45mm×45mm, the wave profile is in-between above two cases. The rarefaction is broadened and its amplitude becomes smaller as the heater size increases. Thus, it is reasonably postulated that the appearance of rarefaction is gradually delayed to infinite time instance and the amplitude of rarefaction approaches asymptotically to zero in the case of the infinite size heater (1D case). Shown in Figure 4 (b) are the profiles of second sound wave at $d_T$=11.0mm. It is seen that the wave profile for the heater size of 45mm×45mm becomes bipolar type as the distance increases. The width of wave profile for the heater size of 25mm×25mm shortens due to the nonlinearity feature of He II and energy conservation as stated above. It is quite intuitive that second sound wave at a distance very close to the heater surface will show 1D characteristic because of its flat geometry, e.g., 2mm/25mm<<1. However, second sound wave still displays 3D characteristic even in flatter geometry, i.e. 2mm/65mm<<1, as shown in Figure 4 (a). The rarefaction portion is indeed detected in 1D experimental geometry, i.e. a channel with the bottom fully bounded by a heater [28], which is caused by superfluid-leak from the contacting circumference between the channel and the heater if it is not well sealed by indium. Such situation is actually frequently encountered in 1D second sound wave experiment. Thus, it can be concluded from above analysis that rigorously planar second sound wave can only appear in the perfect 1D geometry.

One interesting phenomenon is that the diffraction pattern of second sound wave formed on the plane of ABCD is similar to that of the pulsed monochromatic light passing through a square slot on the screen. Shown in Figures 5 (a) and (b) are the diffraction patterns of second sound wave and pulsed monochromatic light, respectively. The pattern intensity of pulsed monochromatic light is normalized. It is seen that both patterns display very similar characteristics, which implies that heat or entropy wave (second sound wave) in He II may diffract in the similar manner as that of light.

B. "Self-focusing" of 3D pulsed second sound wave

One particular phenomenon is that as He II temperature increases to the range where $\varepsilon(T)$ is negative, the transmitting speed of second sound wave along the central axis will be smaller than that off-axis, as shown by eq. (4). This may lead to a so-called "self-focusing" phenomenon suggested theoretically by Nemirovskii [29] and first observed experimentally by L. C. Krysac [15] at temperature close to $T_\lambda$. It was argued that the amplitude of the rarefaction became smaller as He II temperature approached $T_\lambda$, which resulted from the development of 3D pulsed second sound wave into planar wave. However, it is found from the numerical results that the decrease of the rarefaction amplitude is due to the effect of quantized vortices. The concerned problem becomes more complicated when the effect of quantized vortices is taken into account. The initial VLD, which can be regarded as the background of quantized vortices, ranges generally from $1 \times 10^4/cm^2$ to $1 \times 10^6/cm^2$ [30], and the value depends on the thermo-fluidic history of He II. For example, frequent release of heat pulse in He II and larger intensity of heat flux and longer heating duration result in higher initial VLD for the next event. Actually, initial VLD can be treated as an initial parameter (fitting parameter) used in the numerical calculation [10]. In the calculation, some other dissipative processes, such as viscosity in normal fluid, et al., are not taken into account due to their negligible effects compared with the effect of mutual friction associated with quantized vortices in a short time and small distance problem. We did not extend our measurements to higher temperature because the superconductive temperature sensor was not sensitive enough to detect smaller temperature amplitude of second sound wave in the vicinity of $T_\lambda$. And thus the calculation results are compared with the experimental results of Krysac [15].

Shown in Figures 6 (a) and (b) are the calculation results of 3D pulsed second sound waves free of and subject to quantized vortices close to $T_\lambda$ at 2.13K, respectively, when the heat flux is 10.0W/cm$^2$. Evidently, the amplitude of the rarefaction of 3D second sound wave is still large when it is free of quantized vortices; however, the amplitude becomes much smaller when it is subject to quantized vortices. And the present calculation result of the temperature excursion is very close to the experimental result of Krysac [15]. As visualized in 3D space, it is found that second sound wave is still in spherical-like shape, however, the amplitude of the positive temperature excursion becomes smaller and the rarefaction becomes less pronounced, as shown by a 2D slice in Figure 7. A higher temperature region is also seen adjacent to the heater surface, which is attributed to the slow decay of quantized vortices. And a large amount of energy accumulates in this region and is gradually transported in a diffusion-like heat transfer mechanism. The changing from second sound wave mechanism into diffusion-like mechanism is recently shown [31] and the existence of such a higher temperature region in one-dimensional case has been analyzed [18, 20] and experimentally proven [8]. The decay of this higher temperature region is very slow, and the velocity of it is in the magnitude of hundred-mm/s from the numerical results, which is much smaller than the velocity of second sound wave. Although the heat can not be diffused by superfluid since it is inviscid, the mutual friction between the superfluid and normal fluid components mediated by quantized vortices may lead to the diffusion-like heat transfer in He II. It may be concluded that the presence of quantized vortices makes the rarefaction of second sound wave less pronounced and the amplitude of the positive temperature excursion smaller and it seems that dissipation induced by quantized vortices plays an important role in producing such phenomenon. Nevertheless, the profile shape of second sound wave is still 3D, but not 1D (planar wave). The possible reason why Krysac [15] regarded her experimental result as planar wave is that it was only the temperature measurement at one point in He II bath, which could not reflect 3D characteristic.

C. Boiling phenomena

It has been frequently reported that boiling phenomena may be detected in second sound wave experiments in a channel [8, 32], for example, Nemirovskii [32] provided a diagram of transient heat transfer regime of He II in terms of heat flux and heating duration. Actually, boiling phenomena were also observed by using the superconductive temperature sensor at the heat flux of 10.0W/cm$^2$ with a longer heating duration of 5.0ms. Shown in Figure 8 is the experimental temperature excursion detected along the central axis at $d_T$ =2.0mm. There are two large temperature overshoots of about 40.0mK. It is estimated from Clapeyron-Clausius equation as given below that a temperature rise must be met to make the thermodynamic state of He II across the saturated vapor pressure line

$$\Delta T = (\frac{\partial T}{\partial P})_{SVP} \Delta P \qquad (5)$$

Where $\Delta P$ is the pressure head produced from the immersion depth over the heater surface, $(\partial T/\partial P)_{SVP}$ is the derivation of the temperature with pressure along the saturated vapor pressure line. A temperature rise of about 89.55mK is required to make the thermodynamic state of He II across the saturated vapor pressure line at an immersion depth of about 20.0cm at 1.6K in the present case. As can be seen from the figure, the temperature overshoots in Figure 8 are smaller than 89.55mK, thus, they are essentially induced by the propagation of the thermal boundary layers which are formed by a mass of quantized vortices in He II. The contact of He II with higher temperature heater surface by re-condensation of vapor layer induces a new thermal boundary layer which propagates away from the heater surface and is then detected by the superconductive temperature sensor. The propagation velocity of the thermal boundary layer in this case is around 500mm/s, which is in agreement with the magnitude of numerical calculation. Furthermore, the rarefaction portion disappears at longer heating duration as in this case, which can be attributed to the fast development of dense quantized vortices in the vicinity of the heater surface. Consequently, dense quantized vortices form a higher temperature region, i.e. thermal boundary layer and a large amount of heat is accumulated in, which leads to the superfluidity locally being broken down and results in the occurrence of boiling phenomenon.

IV. CONCLUSION

The study of 3D pulsed second sound wave is carried out on the basis of two-fluid model with Vinen's VLD equation. The characteristic feature of 3D second sound wave is a rarefaction following the positive temperature excursion, which has been clearly presented by the numerical analysis. And the numerical result agrees quite well with the experimental measurement. The counterflow between the superfluid and normal fluid components becomes *inverse* in the rarefaction of the running second sound wave (free of reflection). It is demonstrated that heater size has a big influence on the profile of 3D second sound wave and rigorously planar second sound wave can only be detected in perfect 1D geometry. The diffraction pattern of 3D pulsed second sound wave on plane ABCD is similar to that of pulsed monochromatic light. Further study of second sound wave near $T_\lambda$ shows that the rarefaction of 3D pulsed second sound wave becomes less pronounced when it is subject to quantized vortices, which explains the outcome of Krysac's experiment [15]. However, the inspection of the results in 3D space reveals that second sound wave is still a 3D one, but not a planar one. Further increase of heating duration leads to the disappearance of the rarefaction of 3D pulsed second sound wave and boiling of He II.

ACKNOWLEDGEMENT

This research is jointly supported by FANEDD (200236) and National Natural Science Foundation of China (50306014) and NCET. We appreciate Prof. S. K. Nemirovskii for invaluable discussion and comments, and also thanks are given to anonymous referees for many useful comments, which improve the quality of the paper.

APPENDIX A

The boundary conditions needed in the calculation are defined in the following way. To initiate second sound wave, a pulsed heat flux is released in He II from the heater surface, which is formulated as

$$\begin{aligned} q_{input} &= q \quad 0 < t \leq t_h \\ q &= 0 \quad t > t_h \end{aligned} \tag{A1}$$

as stated in the text, the energy is totally borne by the normal fluid component, and thus, yielding $q = \rho s T v_n$. By combining $\rho = \rho_s + \rho_n$ and $\rho v = \rho_s v_s + \rho_n v_n$ and considering the rigid heater surface with $\rho v = 0$, then $v_{ns,\perp} = q / \rho_s s T$ is the boundary condition for normal direction to the heater surface. Due to the fact that heat is uniformly added and the heater is considered to be infinitely thin and with no thermal capacitance, heat can be considered being added only at normal direction, which obviously leads to $v_{ns,//} = 0$.

The reflection of second sound wave at the boundary wall can be treated as full reflection since the wall is adiabatic, and moreover, the slippery boundary condition can be applied because of the negligible viscosity of He II. By taking these points into consideration, the boundary conditions can be considered being the "thermal" mirror conditions, i.e. the imaginary grid points next to the boundary at both sides (left or right of the boundary and below or above the boundary) are symmetric.

For the scalar quantities, such as the thermodynamic quantities and the vortex line density, the boundary conditions can be formulated as $f_{left} = f_{right}$ or $f_{above} = f_{below}$. However, for the vector quantities, such as the velocity $v$ and the mass flux $\rho v$, the boundary conditions have to be dealt with separately. In detail, at the boundaries of DCC'D' and CBB'C', the boundary conditions for the velocity $v$ and the mass flux $\rho v$ can be formulated as

$$v_{\perp,left} = -v_{\perp,right} \text{ and } v_{//,left} = v_{//,right}$$
$$\rho v_{\perp,left} = -\rho v_{\perp,right} \text{ and } \rho v_{//,left} = -\rho v_{//,right} \tag{A2}$$

at the boundaries of DAA'D' and ABB'A', the boundary conditions for the velocity $v$ and the mass flux $\rho v$ are of a little bit difference, since they are essentially the middle planes of the whole domain. And they can be formulated as following due to symmetry

$$v_{\perp,left} = -v_{\perp,right} \text{ and } v_{//,left} = v_{//,right}$$
$$\rho v_{\perp,left} = -\rho v_{\perp,right} \text{ and } \rho v_{//,left} = \rho v_{//,right} \tag{A3}$$

The boundary conditions for A'B'C'D' and ABCD other than the area of the heater surface can be dealt with in the similar manner, and they are formulated as

$$v_{\perp,above} = -v_{\perp,below} \text{ and } v_{//,above} = v_{//,below}$$
$$\rho v_{\perp,above} = -\rho v_{\perp,below} \text{ and } \rho v_{//,above} = -\rho v_{//,below} \tag{A4}$$

APPENDIX B

It has been shown in section III A, that the profile of 3D pulsed second sound drastically influenced by the heater size and it turns into 1D pulsed second sound wave when the bottom of the channel is fully bounded by the heater. It is seen from the calculation results of 1D pulsed second sound wave shown in Figure B1 that the rarefaction disappears in the temperature excursion. Although only the data up to 3.0ms are shown in Figure B1 (a), as can be seen from 2D slice in Figure B1 (b), no rarefaction forms behind the positive second sound wave portion. Thus, it will only display the characteristic of planar second sound wave during propagation.

Kolmakov, E. V. Lebedeva, L. P. Mezhov-Deglin, and A. B. Trusov, JETP Letters, **69**, 767 (1999)

Figure captions:

FIG. 1 The schematic illustration of the calculation domain. The dimension is 45mm(AB) × 45mm(AD) × 90mm(AA'). The solid area shown in the figure represents one fourth of the heater (only 3D case is shown, the detailed description of 1D and 2D cases can be found in the text).

FIG. 2 The comparison of the experimental result with the calculation result of 3D pulsed second sound wave with a heater of 25mm×25mm in size. He II bath temperature $T_b$ =1.6K; heat flux $q$ =10.0W/cm$^2$; heating duration $t_h$ =0.5ms (a) and $t_h$ =1.0ms (b); and the distance $d_T$ between the superconductive temperature sensor and the heater surface along the central axis is 2.0mm. In the analysis, second sound wave is assumed to be free of quantized vortices because the time interval between the releases of the heat pulses into He II is long enough to let quantized vortices to decay to the initial level.

FIG. 3 (Color online) A series of 2D slices of 3D pulsed second sound wave at different time instances. (a) $t$ =0.5ms, (b) $t$ =0.75ms, (c) $t$ =1.0ms, (d) $t$ =2.0ms, (e) the relative velocity distribution at $t$ =2.0ms. The relative velocity ($v_{ns}$) distribution is the projection of the velocity vectors in 3D space along the plane of ABB'A'. He II bath temperature $T_b$ =1.6K; heat flux $q$ =10.0W/cm$^2$; heating duration $t_h$ =0.5ms; the release of heat flux is at $t$ =0ms. The thick black lines in 2D slices indicate the heater size (25mm×25mm) and location.

FIG. 4 (Color online) Profiles and 2D slices of 3D pulsed second sound wave. (a) profiles of 3D pulsed second sound wave with different heater sizes at $d_T$ =2.0mm along the central axis; (b)

profiles of 3D pulsed second sound wave with different heater sizes at $d_T=11.0$mm along the central axis; (c) 2D slice at $t=1.0$ms; (d) 2D slice at $t=2.0$ms; (e) 2D slice at $t=3.0$ms. He II bath temperature $T_b=1.6$K; heat flux $q=10.0$W/cm$^2$; heating duration $t_h=0.5$ms; the thick black lines in 2D slices indicate the heater size (65mm×65mm) and location.

FIG. 5 (Color online) Diffraction pattern of second sound wave on the plane of ABCD (a) and that of pulsed monochromatic light (b). The area designated by the square indicates the heater or the diffraction slot.

FIG. 6 The calculation results of 3D pulsed second sound wave free of (a) and subject to (b) quantized vortices. He II bath temperature $T_b=2.13$K, heat flux $q=10.0$W/cm$^2$, heating duration $t_h=1.0$ms and the temperature excursion in the figure is at the same location as that in Reference [15], which is at 22.0mm above the heater surface and 3.0mm off the central axis. In FIG. 6 (b), initial VLD is $3.0\times10^5$/cm$^2$.

FIG. 7 (Color online) 2D slice of 3D pulsed second sound wave close to $T_\lambda$ at 2.13K at $t=4.0$ms, initial VLD is $3.0\times10^5$/cm$^2$. The higher temperature region adjacent to the heater surface is due to accumulating of dense quantized vortices and quite a large amount of energy is left un-transported and is then gradually transported in a diffusion-like manner. As seen in the figure, although the rarefaction almost disappears, second sound wave is still a non-planar one. The thick black line in 2D slice indicates the heater size (16mm×16mm, same as that in Reference [15]) and location.

FIG. 8  The temperature excursion detected at $d_T=2.0$mm above the heater surface along the central axis in the case of longer heating duration of 5.0ms. The temperature excursion is composed of two parts including second sound wave and thermal boundary layer.

FIG. B1  (Color online) The calculation results of 1D pulsed second sound wave when the heater size is enlarged to bound the bottom (ABCD in Figure 1) of the channel perfectly. (a) The profile of 1D pulsed second sound wave at two locations: A is at $d_T=2.0$mm above the heater surface along the central axis, B is at 11.0mm above the heater surface and 12.0mm off the central axis; (b) 2D slice of second sound wave at $t=2.0$ms. The thick black line in 2D slice indicates the heater size and location.

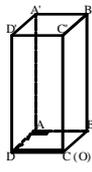

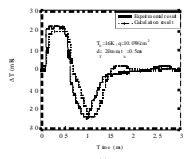

(a)

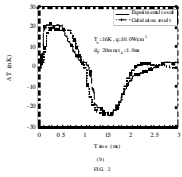

(b)

FIG. 2

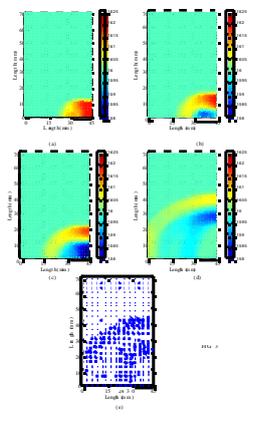

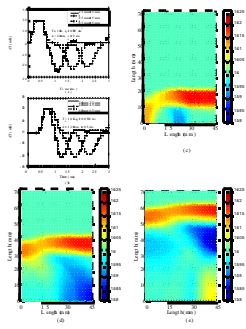

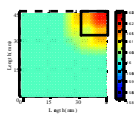

(a)

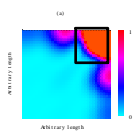

(b)

FIG. 5

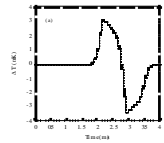
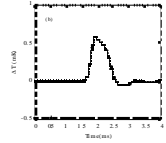

Fig. 6
27

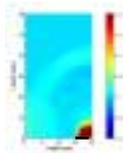

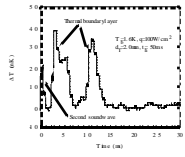

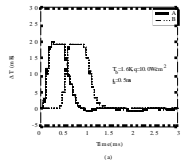

(a)

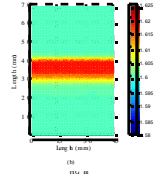

(b)

30